\pdfoutput=1
\documentclass[%
  letterpaper,%
  twocolumn,%
  amsmath,%
  amssymb,%
  amsfonts,%
  superscriptaddress,%
  groupedaddress,% 
  prb%
]{revtex4}
\usepackage[flushleft]{threeparttable}
\usepackage{bm,graphicx,hyperref}
\hypersetup{%
  breaklinks = {true},
  citecolor = {blue},
  colorlinks = {true},
  linkcolor = {red},
}

\begin{document}

\title{Highly Deformable Graphene Kirigami}

\author{Zenan~Qi\footnote{Electronic address: zenanqi@bu.edu}}
\affiliation{Department of Mechanical Engineering, Boston University, Boston, MA 
02215}

\author{David~K.~Campbell\footnote{Electronic address: dkcampbe@bu.edu}}
\email[Corresponding author: ]{dkcampbe@bu.edu}
\affiliation{Department of Physics, Boston University, 590 Commonwealth Ave, 
Boston, Massachusetts 02215, USA}

\author{Harold~S.~Park\footnote{Electronic address: parkhs@bu.edu}}
\email[Corresponding author: ]{parkhs@bu.edu}
\affiliation{Department of Mechanical Engineering, Boston University, Boston, MA 
02215}

\date{\today}

\begin{abstract}

Graphene's exceptional mechanical properties, including its highest-known stiffness (1 TPa) and strength (100 GPa) have been exploited for various structural applications. However, graphene is also known to be quite brittle, with experimentally-measured tensile fracture strains that do not exceed a few percent.  In this work, we introduce the notion of graphene kirigami, where concepts that have been used almost exclusively for macroscale structures are applied to dramatically enhance the stretchability of both zigzag and armchair graphene. Specifically, we show using classical molecular dynamics simulations that the yield and fracture strains of graphene can be enhanced by about a factor of three using kirigami as compared to standard monolayer graphene. 
This enhanced ductility in graphene should open up interesting opportunities not only mechanically, but also in coupling to graphene's electronic behavior.

\end{abstract}

\maketitle

%\section{Introduction}

Despite being a two-dimensional material that is only one plane of atoms thick, monolayer graphene exhibits a very desirable combination of mechanical properties. These include both a high Young's modulus of about 1 TPa, as well as an intrinsic strength of about 100 GPa~\cite{leeSCIENCE2008}, where both of these quantities are about one order of magnitude larger than is observed in commonly used structural materials such as steel. These properties have enabled graphene-based polymer nanocomposites~\cite{pottsPOLYMER2011,ramanNN2008}, stretchable electronics~\cite{kimNATURE2009}, and nanoelectromechanical systems (NEMS) and nanoresonators~\cite{bunchSCIENCE2007}.

While exhibiting high strength and stiffness, graphene's mechanical performance is hindered by its brittle nature, where under tensile loading graphene fractures immediately after yielding at strains generally not exceeding a few percent~\cite{zhaoNL2009,zhaoJAP2010,luMSMSE2011,liuPRB2007}.  A key issue then for graphene is to not only develop techniques to enhance its ductility, but to do so in a systematic, tunable fashion.  One example in this direction is the recent work of Zhu et al.~\cite{ZhuAPL2014}, who found that graphene nanomeshes can be stretched to nearly 50\% strain.  While the nanomeshes do enable substantial increases in mechanical stretchability, there is considerably greater opportunity to tailor the shapes and hence physical properties of graphene using the principles of kirigami, which is a version of origami in which cutting is used to change the morphology of a structure.  Examples of the structural and geometric diversity that can be achieved using kirigami approaches for graphene have already been demonstrated experimentally~\cite{BleesAPS2014}.

Accordingly, we present in this work the result of classical molecular dynamics (MD) simulations on the tensile deformation of a specific, experimentally-realized form of graphene kirigami~\cite{BleesAPS2014}. We demonstrate using MD simulations that the resulting monolayer graphene kirigami can sustain yield and fracture strains that can be more than three times larger than can pristine, bulk graphene. While kirigami has traditionally been applied to increase the flexibility of macroscale structures, here we demonstrate that its benefits extend down to single-layer, two-dimensional nanomaterials.  Finally, we introduce two non-dimensional design constants that we show can be used to tailor and tune the mechanical properties of the kirigami.

%\section{Numerical Results}

%---------------------------------------------------------------------
\begin{figure}
  \centering
  \includegraphics[scale=0.5]{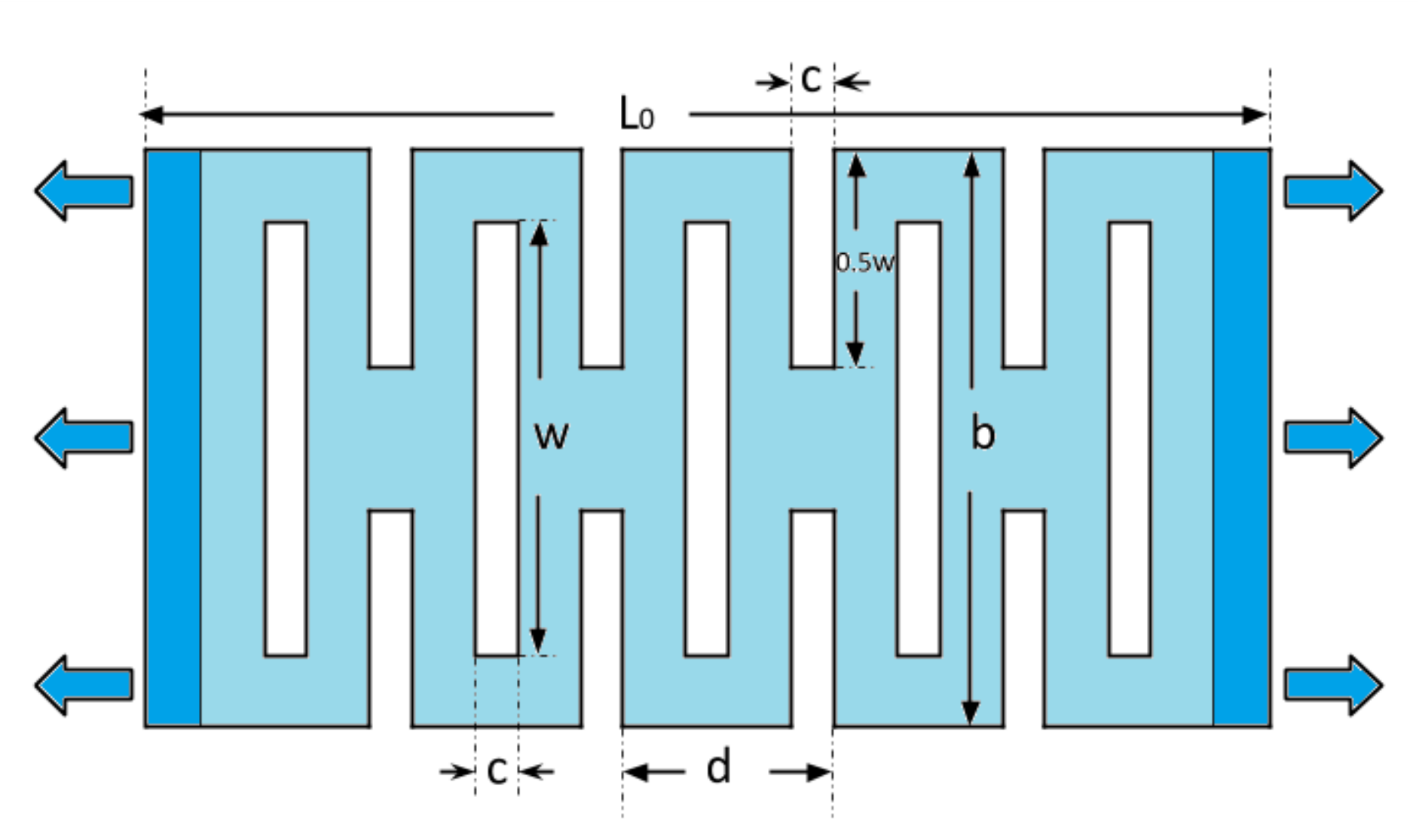}
  \caption{(Color online)  Schematic of the graphene kirigami, with key geometric parameters labeled.  The kirigami is deformed via tensile displacement loading that is applied at the two ends in the direction indicated by the arrows. }
  \label{fig:schematic}
\end{figure}
%---------------------------------------------------------------------

Our MD simulations were done with the Sandia-developed open source code LAMMPS~\cite{plimptonLAMMPS,PlimptonJCP1995}.  We used the AIREBO potential~\cite{stuartJCP2000} to describe the C-C interactions, as this potential has been shown to describe accurately the various carbon interactions including bond breaking and reforming~\cite{zhaoJAP2010,qiNANO2010}.  The cut-off radius for the REBO term is 2~\AA~and the cut-off radius for the Lennard-Jones term in the AIREBO potential is 6.8~\AA. The graphene kirigami were constructed by making cuts in a graphene nanoribbon, with the resulting kirigami shown schematically in Fig.~\ref{fig:schematic}.  

The graphene kirigami in Fig.~\ref{fig:schematic} is marked by several key geometric features, which we now describe.  First, the length of the nanoribbon is $L_{0}$, while the width is $b$.  The height of each interior cut is $w$, while the width of each interior cut is $c$.  The distance between successive kirigami cuts is $d$,  while the edge cut length is defined to be half of the interior cut length (i.e. $0.5w$).  For simplicity, all of the half cut lengths are the same, while all of the interior cut lengths were also fixed.  While the dimensions of the kirigami changed according to the parametric studies we performed, a representative kirigami structure we studied had 11408 atoms, $L_{0}\sim$340~\AA, $w\sim$67~\AA, $b\sim$100~\AA, $c\sim$5~\AA~and successive kirigami cut distance of $d\sim$48~\AA. Our discussion below on the deformation mechanisms and failure process will be based on this specific geometry, though we will report trends in mechanical properties based on a range of geometric parameters, as we will describe later.

%---------------------------------------------------------------------
\begin{figure*}
  \centering
  \includegraphics[width=0.99\textwidth]{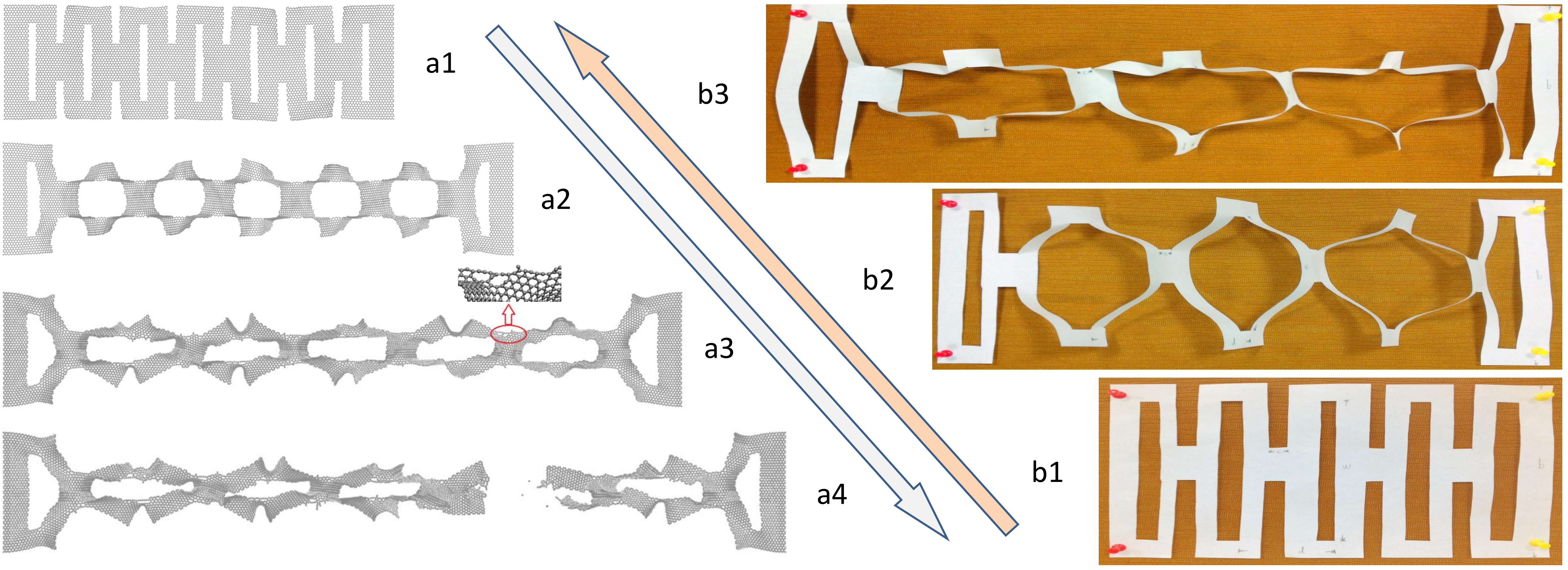}
  \caption{(Color online)  Left column: snapshots of the top view (a1-a4) illustrating the deformation stages for zigzag graphene kirigami.  A representative yield region is marked in (a3). The tensile strains corresponding to the different stages are 14\%, 29\%, 56\% and 65\% respectively. Right column: schematic top view pictures (b1-b3) of similarly patterned paper kirigami for comparison. (b1-b3) correspond to (a1-a3) while paper kirigami fracture picture is not shown. Graphene figures were generated by VMD~\cite{Humphrey1996}. \it{All snapshots were scaled for purposes of simplicity of visualization}.} 
  \label{fig:snapshots}
\end{figure*}
%---------------------------------------------------------------------

\begin{figure}
  \centering
  \includegraphics[scale=0.6]{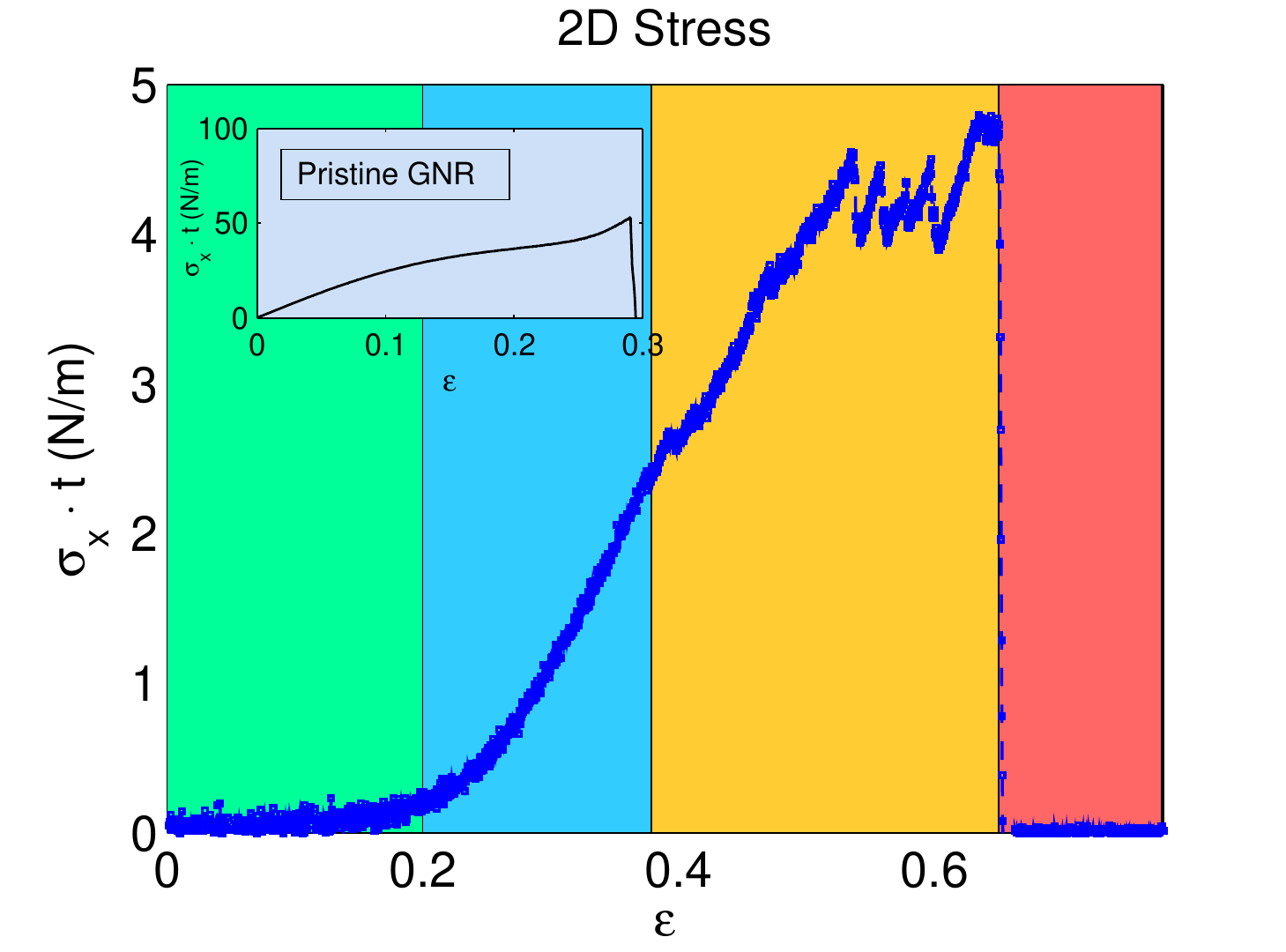}
  \caption{(Color online)  Stress-strain curve of the representative graphene kirigami as shown in Fig.~\ref{fig:snapshots}, where the 2D stress was calculated as stress $\sigma$ times thickness $t$. Green, blue, orange and red regions correspond to the four stages of deformation discussed in the text and illustrated in Fig.~\ref{fig:snapshots}.  A stress-strain curve of a pristine zigzag graphene nanoribbon with the same width is shown in the inset for comparison.}
  \label{fig:stress-strain}
\end{figure}

The kirigami structure was first relaxed for 10 ps within the constant temperature (NVT) ensemble at room temperature (300K). We primarily considered zigzag chirality, though simulations of armchair graphene were also conducted to verify that the results we present are qualitatively independent of chirality. Non-periodic boundary conditions were used in three directions.  The kirigami was deformed in tension within the same NVT ensemble by applying a uniform displacement loading on both edges, resulting in a strain rate of $\sim 10^9$ s$^{-1}$ until fracture occurred.  To illustrate the deformation response, we show a series of snapshots of the representative stages during elongation in Fig.~\ref{fig:snapshots} along with the tensile stress strain curve in Fig.~\ref{fig:stress-strain} for the zigzag graphene kirigami configuration illustrated in Fig.~\ref{fig:schematic}.

Fig.~\ref{fig:snapshots}(a1)-(a4) shows that the graphene kirigami exhibits four distinct stages preceding fracture.  Before any tensile loading is applied, the structure ripples out of plane during the initial thermal equilibration stage.  Once tensile loading is applied, as shown in (a1), the kirigami structure elongates, with the interior cuts exhibiting tensile elongations of roughly 20\% strain along the loading direction.  While the interior cuts are initially vertical after the thermal equilibration, during this initial stage of tensile loading (for strains smaller than about 20\%), the cuts flip and rotate such that they make a nearly 45 degree angle with the loading direction, as shown in (a1).  This flipping and rotation is the key mechanism that enables the high ductility of graphene kirigami, and during this stage kirigami structure is elongated without significantly stretching carbon bonds. This can be seen from Fig.~\ref{fig:stress-strain} where the stress is nearly zero in this stage (green region). In the second stage, shown in (a2), the carbon bonds start to be stretched together with the strained kirigami structure causing the stress increase as shown in Fig.~\ref{fig:stress-strain} for strains between about 20 and 38\% (blue region).  We note that the deformation in the first two stages, which accounts for nearly 40\% tensile strain, is elastic and reversible.

%---------------------------------------------------------------------
%---------------------------------------------------------------------

Yielding begins in the third stage, as shown in (a3) at a global tensile strain of almost 40\%.  The yielding initiates from the tips of the interior cuts, as marked in (a3), as those tips exhibit high stress concentrations due to the large deformations.  Finally, fracture occurs in the fourth stage at a strain of about 65\% in (a4). We also studied armchair graphene kirigami structures and found similar deformation patterns.  We note that (a2) are snapshots before yield while (a3) are after yield in Fig.~\ref{fig:snapshots}.  

To demonstrate that the atomic scale, single-layer graphene kirigami deforms similarly to macroscale kirigami, we created paper kirigami using A4 paper with similar geometric parameters, and subjected it to uniaxial stretching as shown in Fig. \ref{fig:snapshots}(b1-b3).  As can be seen, the graphene and paper kirigami exhibit qualitatively similar deformation features, which shows that many of the known advantages of macroscale kirigami may hold even for a single-layer, two-dimensional material. We note that fracture of the paper kirigami is not shown for preservation purposes.

Having established that kirigami is an effective method to enhance stretchability in graphene, one key challenge is to systematically understand how the geometric parameters of the kirigami shown in Fig.~\ref{fig:schematic} impact the key mechanical properties of interest, {\it i.e.} the yield stress and strain, as well as the fracture strain. Such an understanding will enable experimentalists to design graphene kirigami that possesses a desired combination of mechanical properties.  To this end, we define two dimensionless parameters that characterize the mechanical properties of the kirigami:  $\alpha=(w-0.5b)/L_0$ and $\beta = (0.5d-c)/L_0$.  Apparently, the number of cuts will directly affect the mechanical response of the kirigami, and thus these parameter choices are based on the assumption that all cases contain the same number of cuts, namely seven middle cuts and six edge cut pairs for all the cases studied in this paper as shown in Fig.~\ref{fig:snapshots}.  Verification of the choices for $\alpha$ and $\beta$ as the appropriate geometric parameters is given in the Supplementary Material.

%---------------------------------------------------------------------
\begin{figure}
  \centering
  \includegraphics[scale=0.6]{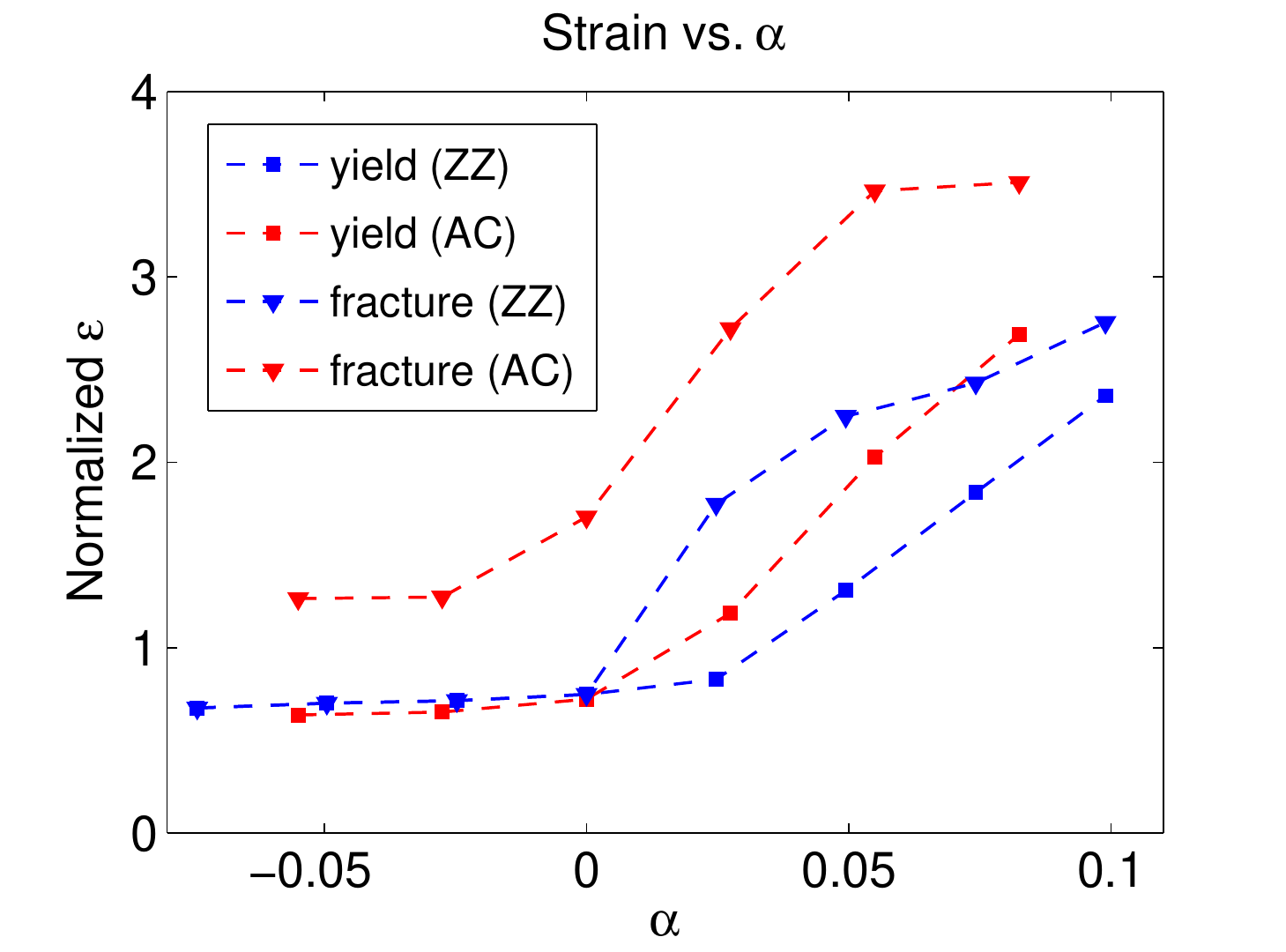}
  \caption{(Color online)  Influence of $\alpha$ on yield strain and fracture strain for zigzag (ZZ) and armchair (AC) graphene kirigami, for constant $\beta=0.057$. Data are normalized by  graphene nanoribbon results with the same width.}
  \label{fig:alpha}
\end{figure}
%---------------------------------------------------------------------

The first parameter, $\alpha$, is the ratio of the overlapping cut length to the nanoribbon length, and controls how much the interior cut, and thus the kirigami, can elongate during tensile deformation.  Specifically, $\alpha$ affects the yield strain and fracture strain due to the flipping elongation mechanism shown in Fig.~\ref{fig:snapshots}(a2).  The yield strain and fracture strain for different values of $\alpha$ are shown in Fig.~\ref{fig:alpha}. It is clear that for $\alpha>0$, the kirigami becomes significantly more ductile, where the fracture strain $\epsilon_{frac}$ is normalized by the fracture strain for bulk graphene.  This is because $\alpha = 0$ corresponds to the configuration when the edge cuts and interior cuts just overlap. When $\alpha < 0$, the edge and interior cuts do not overlap and the flipping and rotation mechanism of Fig.~\ref{fig:snapshots}(a2) and (b2) does not occur. In contrast, when $\alpha > 0$, the flip-rotation mechanism for the interior cuts does occur, and enables the kirigami to expand without substantial stretching of the carbon bonds. This is also reflected from the 2D stress-strain curve as shown in Fig.~\ref{fig:stress-strain}, where the stress was calculated as stress times thickness to avoid known controversies in defining the thickness for carbon-based nanostructures~\cite{huangPRB2006}.  For completeness, we note that for the paper kirigami seen in Fig. \ref{fig:snapshots}(b1-b3), the non-dimensional values are $\alpha\sim0.13$ and $\beta\sim0.06$.

The deformation illustrated in Fig.~\ref{fig:snapshots}(a1) and (b1) corresponds to the green region in Fig.~\ref{fig:stress-strain}, where before roughly $\epsilon = 0.2$ the kirigami structure elongates without significant stretching of the carbon bonds, which explains the very low value of stress for that strain region. However, between strains of $\epsilon = 0.2$ to $\epsilon = 0.38$ (yield strain), the carbon bonds begin to be stretched substantially, leading to the increase in stress seen in Fig.~\ref{fig:stress-strain}.  With a further increase in strain, yielding occurs via local fracture of graphene as shown in Fig.~\ref{fig:snapshots}(a3) and (b3).  Eventually, the local fracture propagates and results in global fracture at $\epsilon = 0.65$ as shown in Fig.~\ref{fig:snapshots}(a4) and (b4) and the red region in Fig.~\ref{fig:stress-strain}. 

In contrast to the pristine graphene nanoribbon as shown in Fig. \ref{fig:stress-strain}, it is clear that the stress that can be sustained by the kirigami is about one order of magnitude smaller.  However, the stretchability, as defined by the fracture strain, is increased by more than a factor of two.  Furthermore, the ductility, defined as the strain after yield, is significantly higher for the kirigami, as it can sustain more than 20\% elongation after yield, while the pristine graphene nanoribbon fractures immediately after yielding.

%---------------------------------------------------------------------
\begin{figure}
  \centering
  \includegraphics[scale=0.6]{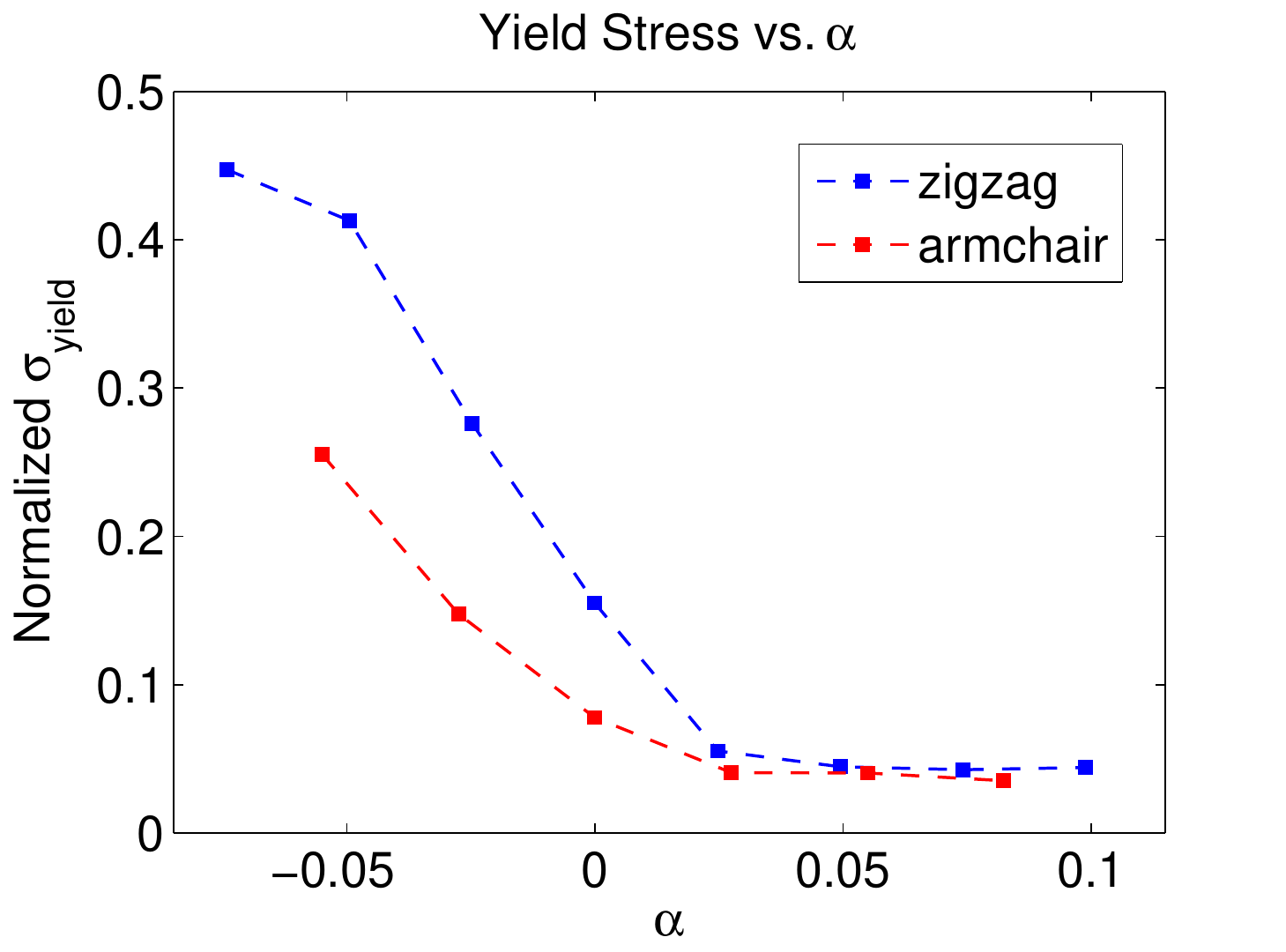}
  \caption{(Color online)  Influence of $\alpha$ on the kirigami yield stress, for constant $\beta=0.057$. Data are normalized by graphene nanoribbon results with the same width.}
  \label{fig:yieldstress}
\end{figure}
%---------------------------------------------------------------------

%---------------------------------------------------------------------
\begin{figure*}
  \centering
  \includegraphics[width=0.9\textwidth]{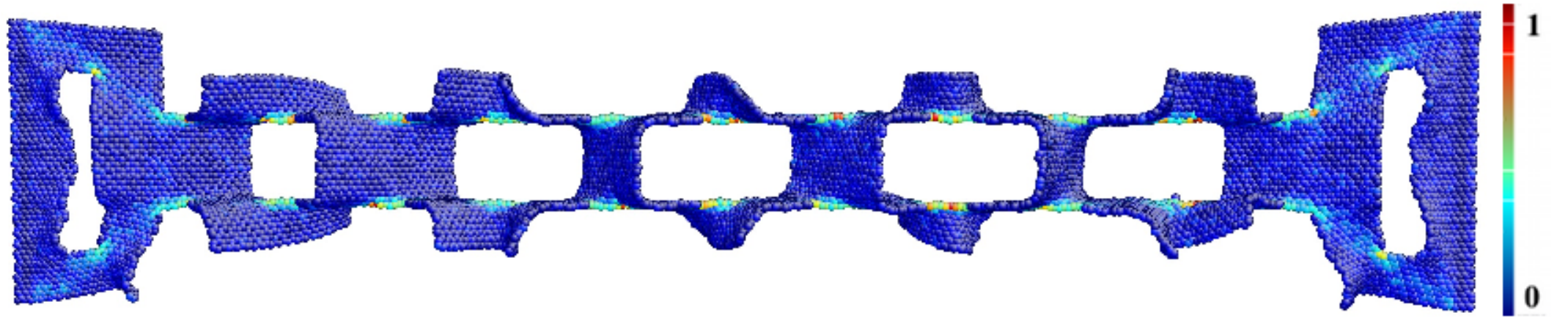}
  \caption{(Color online)  Von Mises stress distribution of zigzag graphene kirigami corresponding to the snapshots in Fig.~\ref{fig:snapshots}(a3), where the data was scaled between 0 to 1. Figure was generated by AtomEye~\cite{LiMSMSE2003}.}
  \label{fig:stressdist}
\end{figure*}
%---------------------------------------------------------------------

While the yield strain increases for increasing $\alpha$, the opposite trend is observed for the yield stress, as shown in Fig.~\ref{fig:yieldstress}. This is also because for negative $\alpha$, the middle and edge cuts do not overlap, and thus the kirigami behaves like a cut-free nanoribbon.  However, when $\alpha$ is positive, the kirigami deforms like the snapshots shown in Fig.~\ref{fig:snapshots} and yields due to the tearing mechanism previously described, where the stress distribution prior to yielding is shown in Fig.~\ref{fig:stressdist}. 

The graphene kirigami thus fractures quite differently as compared to bulk graphene or a graphene nanoribbon.  Instead of brittle fracture, the yielding of graphene kirigami begins from the corners of the interior cuts, and gradually propagates until fracture occurs.  The stress distribution in Fig.~\ref{fig:stressdist} shows that the stress is concentrated at the corners of the interior cuts while being very small in other regions of the kirigami.  This stress localization explains why the yield stress curve turns flat after $\alpha$ becomes positive, as shown in Fig.~\ref{fig:yieldstress}.

The results shown in Figs.~\ref{fig:alpha} and ~\ref{fig:yieldstress} were carried out at a constant value of $\beta = 0.057$. While $\alpha$ describes the geometry perpendicular to the tensile loading direction, $\beta$ describes the geometry parallel to the tensile loading direction. Referring to the kirigami schematic in Fig.~\ref{fig:schematic}, we see that $\beta$ represents the ratio of overlapping width to the nanoribbon length, which is directly related to the density of cuts, and where theoretically $\beta$ can take values ranging from nearly zero to one.  However, in practice, when $\beta$ exceeds about 0.125, the edge atoms between adjacent edge cuts interact and thus break the kirigami structure.  

\begin{figure}
  \centering
  \includegraphics[scale=0.6]{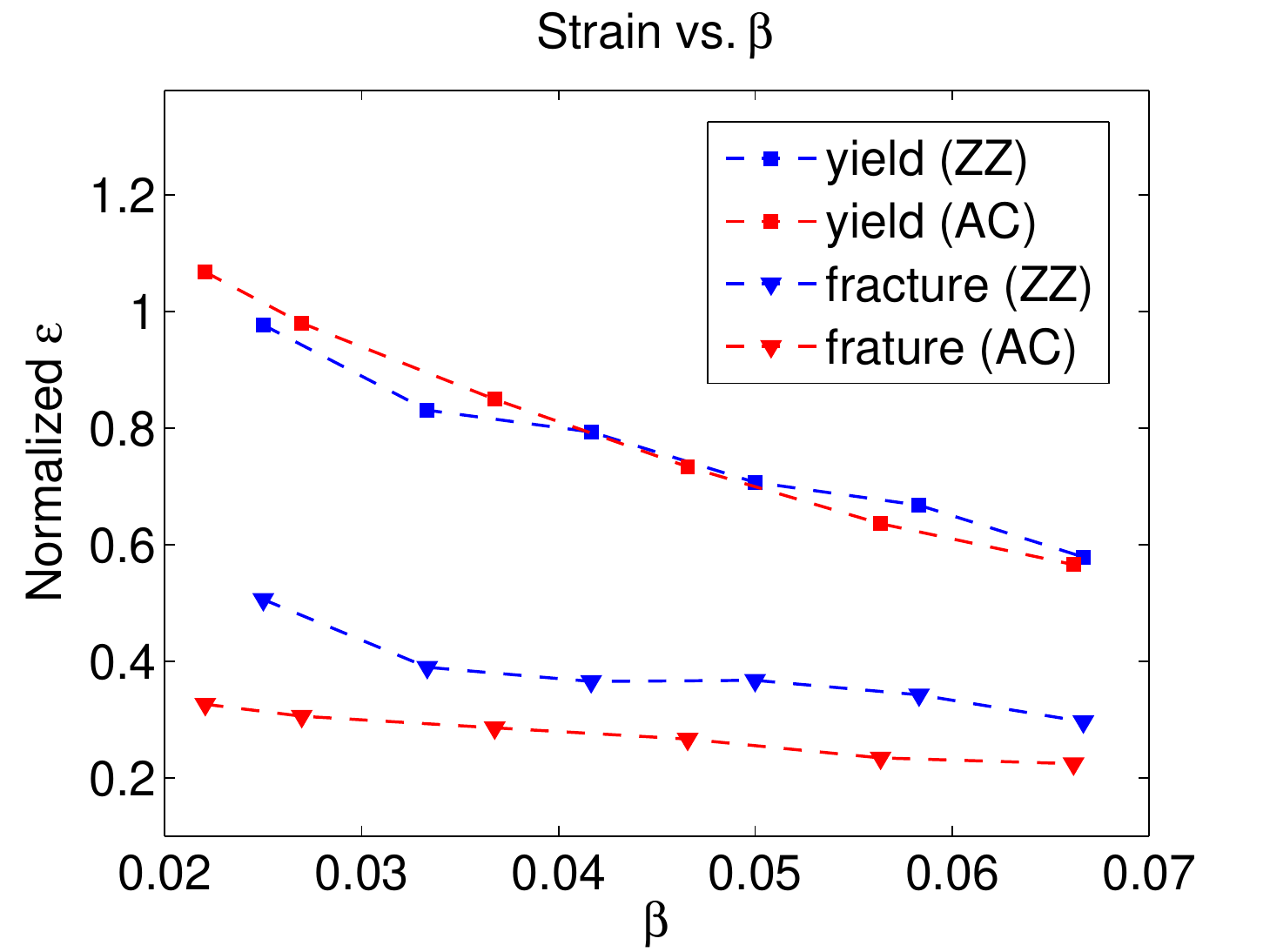}
  \caption{(Color online)  Influence of $\beta$ on fracture strain and yield strain, for constant $\alpha=0.07$. Data are normalized by graphene nanoribbon results with the same width.}
  \label{fig:beta}
\end{figure}

The impact of $\beta$ on the yield strain is shown in Fig.~\ref{fig:beta} for constant $\alpha = 0.07$. In contrast to $\alpha$, which describes the length of the overlapping region, $\beta$ describes the width of the cuts. Furthermore, while the cut length determines how much the kirigami can elongate in along the loading direction, as previously illustrated in Fig.~\ref{fig:alpha}, the cut width determines the aspect ratio of the overlapping region, which controls the likelihood of the flipping and rotating mechanism previously discussed.  Therefore, Fig.~\ref{fig:beta} demonstrates that when $\beta$ increases, the overlapping region width increases, which results in increased difficulty for the flipping and rotation mechanism to occur, resulting in a decrease in the yield and fracture strains.

%---------------------------------------------------------------------
%---------------------------------------------------------------------

Other than the yield strain, fracture strain and yield stress, we also studied the Young's modulus and toughness for the graphene kirigami. Results for the kirigami geometry in Fig.~\ref{fig:schematic} are listed in Table.~\ref{table:ET}, where the Young's modulus was obtained through linear fitting of the stress-strain curve, while the toughness $U_{T}$ was calculated as $U_T = \int_0^{\epsilon_f} \sigma d\epsilon$.  The nature of the stress-strain curve of Fig.~\ref{fig:stress-strain} leads us to define two Young's moduli.  The first ($E_{1}$) corresponds to the low stress region for strains smaller than $\epsilon=0.2$ (green) in Fig.~\ref{fig:stress-strain}, while the second ($E_{2}$) corresponds to the increasing stress region between $\epsilon=0.2$ and $\epsilon=0.38$ (blue).  Table.~\ref{table:ET} illustrates that for both armchair and zigzag graphene kirigami, the Young's modulus and toughness are significantly lower compared to either bulk graphene or graphene nanoribbons. Furthermore, though the kirigami structure significantly enhances the yield and fracture strains for graphene, the order of magnitude reduction in yield stress results in an overall decrease in toughness for graphene kirigami as compared to standard graphene.

\begin{table}
\begin{threeparttable}
\caption{Young's Modulus (E) and Toughness ($U_T$)}
\centering
\begin{tabular}{c c c c} 
\hline\hline 
Case \hspace{1pc} & $E_1$ (N/m) \hspace{1pc} & $E_2$ (N/m) \hspace{1pc} & $U_T$ (N/m) \\ [0.5ex]
\hline
kirigami(ZZ) \hspace{1pc} & 0.80 \hspace{1pc} & 15.17 \hspace{1pc} & 1.21 \\
kirigami(AC) \hspace{1pc} & 0.36 \hspace{1pc} & 11.03 \hspace{1pc} & 1.77 \\ 
nanoribbon(ZZ) \hspace{1pc} & - \hspace{1pc} & 295.91 \hspace{1pc} & 8.27 \\ 
nanoribbon(AC) \hspace{1pc} & - \hspace{1pc} & 304.70 \hspace{1pc} & 4.78 \\ 
bulk(ZZ) \hspace{1pc} & - \hspace{1pc} & 315.53 \hspace{1pc} & 9.29 \\ 
bulk(AC) \hspace{1pc} & - \hspace{1pc} & 319.69 \hspace{1pc} & 5.34 \\ 
\hline
\end{tabular} 
\begin{tablenotes}
      \small
      \item Note: For kirigami (ZZ) case, $L_0 = 340\AA$, $b = 100\AA$, $\alpha \sim 0.05$ and $\beta \sim 0.06$; for kirigami (AC) caes, $L_0 = 347\AA$, $b = 117\AA$, $\alpha \sim 0.05$ and $\beta \sim 0.05$.
\end{tablenotes}
\label{table:ET}
\end{threeparttable}
\end{table} 

%\section{Conclusion}

In summary, we have utilized classical molecular dynamics to perform a systematic study of an experimentally-realized form of graphene kirigami.  In doing so, we have identified two key geometric parameters that can be tuned to controllably and predictably tailor the mechanical properties of graphene kirigami.  Of particular interest, the kirigami structures were found to exhibit yield and fracture strains that can be more than three times that of bulk graphene or graphene nanoribbons.  These simulations demonstrate that the benefits of kirigami patterning, which have been exploited for macroscale structures, may also hold in the thinnest possible nanostructures.  We therefore expect that these kirigami structures may prove to be extremely useful in ameliorating the known brittle behavior of graphene nanostructures, and to provide new methods for producing novel strain engineered graphene devices.

%\acknowledgments
%%DKC: Do we have a anyone to acknowledge except the funding agencies?
HSP and ZQ acknowledge support from the Mechanical Engineering and Physics Departments at Boston University.  DKC acknowledges the hospitality of the International Institute of Physics at the Universidade Federal do Rio Grande do Norte in Natal, Brazil, where part of this work was conducted.

\bibliography{biball}

\begin{thebibliography}{18}
\expandafter\ifx\csname natexlab\endcsname\relax\def\natexlab#1{#1}\fi
\expandafter\ifx\csname bibnamefont\endcsname\relax
  \def\bibnamefont#1{#1}\fi
\expandafter\ifx\csname bibfnamefont\endcsname\relax
  \def\bibfnamefont#1{#1}\fi
\expandafter\ifx\csname citenamefont\endcsname\relax
  \def\citenamefont#1{#1}\fi
\expandafter\ifx\csname url\endcsname\relax
  \def\url#1{\texttt{#1}}\fi
\expandafter\ifx\csname urlprefix\endcsname\relax\def\urlprefix{URL }\fi
\providecommand{\bibinfo}[2]{#2}
\providecommand{\eprint}[2][]{\url{#2}}

\bibitem[{\citenamefont{Lee et~al.}(2008)\citenamefont{Lee, Wei, Kysar, and
  Hone}}]{leeSCIENCE2008}
\bibinfo{author}{\bibfnamefont{C.}~\bibnamefont{Lee}},
  \bibinfo{author}{\bibfnamefont{X.}~\bibnamefont{Wei}},
  \bibinfo{author}{\bibfnamefont{J.~W.} \bibnamefont{Kysar}}, \bibnamefont{and}
  \bibinfo{author}{\bibfnamefont{J.}~\bibnamefont{Hone}},
  \bibinfo{journal}{Science} \textbf{\bibinfo{volume}{321}},
  \bibinfo{pages}{385} (\bibinfo{year}{2008}).

\bibitem[{\citenamefont{Potts et~al.}(2011)\citenamefont{Potts, Dreyer,
  Bielawski, and Ruoff}}]{pottsPOLYMER2011}
\bibinfo{author}{\bibfnamefont{J.~R.} \bibnamefont{Potts}},
  \bibinfo{author}{\bibfnamefont{D.~R.} \bibnamefont{Dreyer}},
  \bibinfo{author}{\bibfnamefont{C.~W.} \bibnamefont{Bielawski}},
  \bibnamefont{and} \bibinfo{author}{\bibfnamefont{R.~S.} \bibnamefont{Ruoff}},
  \bibinfo{journal}{Polymer} \textbf{\bibinfo{volume}{52}}, \bibinfo{pages}{5}
  (\bibinfo{year}{2011}).

\bibitem[{\citenamefont{Ramanathan et~al.}(2008)\citenamefont{Ramanathan,
  Abdala, Stankovich, Dikin, Herrera-Alonso, Piner, Adamson, Schniepp, Chen,
  Ruoff et~al.}}]{ramanNN2008}
\bibinfo{author}{\bibfnamefont{T.}~\bibnamefont{Ramanathan}},
  \bibinfo{author}{\bibfnamefont{A.~A.} \bibnamefont{Abdala}},
  \bibinfo{author}{\bibfnamefont{S.}~\bibnamefont{Stankovich}},
  \bibinfo{author}{\bibfnamefont{D.~A.} \bibnamefont{Dikin}},
  \bibinfo{author}{\bibfnamefont{M.}~\bibnamefont{Herrera-Alonso}},
  \bibinfo{author}{\bibfnamefont{R.~D.} \bibnamefont{Piner}},
  \bibinfo{author}{\bibfnamefont{D.~H.} \bibnamefont{Adamson}},
  \bibinfo{author}{\bibfnamefont{H.~C.} \bibnamefont{Schniepp}},
  \bibinfo{author}{\bibfnamefont{X.}~\bibnamefont{Chen}},
  \bibinfo{author}{\bibfnamefont{R.~S.} \bibnamefont{Ruoff}},
  \bibnamefont{et~al.}, \bibinfo{journal}{Nature Nanotechnology}
  \textbf{\bibinfo{volume}{3}}, \bibinfo{pages}{327} (\bibinfo{year}{2008}).

\bibitem[{\citenamefont{Kim et~al.}(2009)\citenamefont{Kim, Zhao, Jang, Lee,
  Kim, Kim, Ahn, Kim, Choi, and Hong}}]{kimNATURE2009}
\bibinfo{author}{\bibfnamefont{K.~S.} \bibnamefont{Kim}},
  \bibinfo{author}{\bibfnamefont{Y.}~\bibnamefont{Zhao}},
  \bibinfo{author}{\bibfnamefont{H.}~\bibnamefont{Jang}},
  \bibinfo{author}{\bibfnamefont{S.~Y.} \bibnamefont{Lee}},
  \bibinfo{author}{\bibfnamefont{J.~M.} \bibnamefont{Kim}},
  \bibinfo{author}{\bibfnamefont{K.~S.} \bibnamefont{Kim}},
  \bibinfo{author}{\bibfnamefont{J.-H.} \bibnamefont{Ahn}},
  \bibinfo{author}{\bibfnamefont{P.}~\bibnamefont{Kim}},
  \bibinfo{author}{\bibfnamefont{J.-Y.} \bibnamefont{Choi}}, \bibnamefont{and}
  \bibinfo{author}{\bibfnamefont{B.~H.} \bibnamefont{Hong}},
  \bibinfo{journal}{Nature} \textbf{\bibinfo{volume}{457}},
  \bibinfo{pages}{706} (\bibinfo{year}{2009}).

\bibitem[{\citenamefont{Bunch et~al.}(2007)\citenamefont{Bunch, van~der Zande,
  Verbridge, Frank, Tanenbaum, Parpia, Craighead, and
  McEuen}}]{bunchSCIENCE2007}
\bibinfo{author}{\bibfnamefont{J.~S.} \bibnamefont{Bunch}},
  \bibinfo{author}{\bibfnamefont{A.~M.} \bibnamefont{van~der Zande}},
  \bibinfo{author}{\bibfnamefont{S.~S.} \bibnamefont{Verbridge}},
  \bibinfo{author}{\bibfnamefont{I.~W.} \bibnamefont{Frank}},
  \bibinfo{author}{\bibfnamefont{D.~M.} \bibnamefont{Tanenbaum}},
  \bibinfo{author}{\bibfnamefont{J.~M.} \bibnamefont{Parpia}},
  \bibinfo{author}{\bibfnamefont{H.~G.} \bibnamefont{Craighead}},
  \bibnamefont{and} \bibinfo{author}{\bibfnamefont{P.~L.}
  \bibnamefont{McEuen}}, \bibinfo{journal}{Science}
  \textbf{\bibinfo{volume}{315}}, \bibinfo{pages}{490} (\bibinfo{year}{2007}).

\bibitem[{\citenamefont{Zhao et~al.}(2009)\citenamefont{Zhao, Min, and
  Aluru}}]{zhaoNL2009}
\bibinfo{author}{\bibfnamefont{H.}~\bibnamefont{Zhao}},
  \bibinfo{author}{\bibfnamefont{K.}~\bibnamefont{Min}}, \bibnamefont{and}
  \bibinfo{author}{\bibfnamefont{N.~R.} \bibnamefont{Aluru}},
  \bibinfo{journal}{Nano Letters} \textbf{\bibinfo{volume}{9}},
  \bibinfo{pages}{3012} (\bibinfo{year}{2009}).

\bibitem[{\citenamefont{Zhao and Aluru}(2010)}]{zhaoJAP2010}
\bibinfo{author}{\bibfnamefont{H.}~\bibnamefont{Zhao}} \bibnamefont{and}
  \bibinfo{author}{\bibfnamefont{N.~R.} \bibnamefont{Aluru}},
  \bibinfo{journal}{Journal of Applied Physics} \textbf{\bibinfo{volume}{108}},
  \bibinfo{pages}{064321} (\bibinfo{year}{2010}).

\bibitem[{\citenamefont{Lu et~al.}(2011)\citenamefont{Lu, Gao, and
  Huang}}]{luMSMSE2011}
\bibinfo{author}{\bibfnamefont{Q.}~\bibnamefont{Lu}},
  \bibinfo{author}{\bibfnamefont{W.}~\bibnamefont{Gao}}, \bibnamefont{and}
  \bibinfo{author}{\bibfnamefont{R.}~\bibnamefont{Huang}},
  \bibinfo{journal}{Modelling and Simulation in Materials Science and
  Engineering} \textbf{\bibinfo{volume}{19}}, \bibinfo{pages}{054006}
  (\bibinfo{year}{2011}).

\bibitem[{\citenamefont{Liu et~al.}(2007)\citenamefont{Liu, Ming, and
  Li}}]{liuPRB2007}
\bibinfo{author}{\bibfnamefont{F.}~\bibnamefont{Liu}},
  \bibinfo{author}{\bibfnamefont{P.}~\bibnamefont{Ming}}, \bibnamefont{and}
  \bibinfo{author}{\bibfnamefont{J.}~\bibnamefont{Li}},
  \bibinfo{journal}{Physical Review B} \textbf{\bibinfo{volume}{76}},
  \bibinfo{pages}{064120} (\bibinfo{year}{2007}).

\bibitem[{\citenamefont{Zhu et~al.}(2014)\citenamefont{Zhu, Huang, and
  Li}}]{ZhuAPL2014}
\bibinfo{author}{\bibfnamefont{S.}~\bibnamefont{Zhu}},
  \bibinfo{author}{\bibfnamefont{Y.}~\bibnamefont{Huang}}, \bibnamefont{and}
  \bibinfo{author}{\bibfnamefont{T.}~\bibnamefont{Li}}, \bibinfo{journal}{apl}
  \textbf{\bibinfo{volume}{104}}, \bibinfo{eid}{173103} (\bibinfo{year}{2014}).

\bibitem[{Ble()}]{BleesAPS2014}
\bibinfo{howpublished}{http://meetings.aps.org/link/BAPS.2014.MAR.L30.11}.

\bibitem[{\citenamefont{Lammps}(2012)}]{plimptonLAMMPS}
\bibinfo{author}{\bibnamefont{Lammps}},
  \bibinfo{journal}{http://lammps.sandia.gov}  (\bibinfo{year}{2012}).

\bibitem[{\citenamefont{Plimpton}(1995)}]{PlimptonJCP1995}
\bibinfo{author}{\bibfnamefont{S.}~\bibnamefont{Plimpton}},
  \bibinfo{journal}{Journal of Computational Physics}
  \textbf{\bibinfo{volume}{117}}, \bibinfo{pages}{1} (\bibinfo{year}{1995}).

\bibitem[{\citenamefont{Stuart et~al.}(2000)\citenamefont{Stuart, Tutein, and
  Harrison}}]{stuartJCP2000}
\bibinfo{author}{\bibfnamefont{S.~J.} \bibnamefont{Stuart}},
  \bibinfo{author}{\bibfnamefont{A.~B.} \bibnamefont{Tutein}},
  \bibnamefont{and} \bibinfo{author}{\bibfnamefont{J.~A.}
  \bibnamefont{Harrison}}, \bibinfo{journal}{Journal of Chemical Physics}
  \textbf{\bibinfo{volume}{112}}, \bibinfo{pages}{6472} (\bibinfo{year}{2000}).

\bibitem[{\citenamefont{Qi et~al.}(2010)\citenamefont{Qi, Zhao, Zhou, Sun,
  Park, and Wu}}]{qiNANO2010}
\bibinfo{author}{\bibfnamefont{Z.}~\bibnamefont{Qi}},
  \bibinfo{author}{\bibfnamefont{F.}~\bibnamefont{Zhao}},
  \bibinfo{author}{\bibfnamefont{X.}~\bibnamefont{Zhou}},
  \bibinfo{author}{\bibfnamefont{Z.}~\bibnamefont{Sun}},
  \bibinfo{author}{\bibfnamefont{H.~S.} \bibnamefont{Park}}, \bibnamefont{and}
  \bibinfo{author}{\bibfnamefont{H.}~\bibnamefont{Wu}},
  \bibinfo{journal}{Nanotechnology} \textbf{\bibinfo{volume}{21}},
  \bibinfo{pages}{265702} (\bibinfo{year}{2010}).

\bibitem[{\citenamefont{Humphrey et~al.}(1996)\citenamefont{Humphrey, Dalke,
  and Schulten}}]{Humphrey1996}
\bibinfo{author}{\bibfnamefont{W.}~\bibnamefont{Humphrey}},
  \bibinfo{author}{\bibfnamefont{A.}~\bibnamefont{Dalke}}, \bibnamefont{and}
  \bibinfo{author}{\bibfnamefont{K.}~\bibnamefont{Schulten}},
  \bibinfo{journal}{Journal of Molecular Graphics}
  \textbf{\bibinfo{volume}{14}}, \bibinfo{pages}{33} (\bibinfo{year}{1996}).

\bibitem[{\citenamefont{Huang et~al.}(2006)\citenamefont{Huang, Wu, and
  Hwang}}]{huangPRB2006}
\bibinfo{author}{\bibfnamefont{Y.}~\bibnamefont{Huang}},
  \bibinfo{author}{\bibfnamefont{J.}~\bibnamefont{Wu}}, \bibnamefont{and}
  \bibinfo{author}{\bibfnamefont{K.~C.} \bibnamefont{Hwang}},
  \bibinfo{journal}{Physical Review B} \textbf{\bibinfo{volume}{74}},
  \bibinfo{pages}{245413} (\bibinfo{year}{2006}).

\bibitem[{\citenamefont{Li}(2003)}]{LiMSMSE2003}
\bibinfo{author}{\bibfnamefont{J.}~\bibnamefont{Li}},
  \bibinfo{journal}{Modelling and Simulation in Materials Science and
  Engineering} \textbf{\bibinfo{volume}{11}}, \bibinfo{pages}{173}
  (\bibinfo{year}{2003}).

\end{thebibliography}

\end{document}